\documentclass[apj]{emulateapj}

\usepackage{longtable}
\usepackage{setspace}

\def\cm{{\rm\thinspace cm}}

\def\erg{{\rm\thinspace erg}}

\def\km{{\rm\thinspace km}}

\def\Mpc{{\rm\thinspace Mpc}}

\def\s{{\rm\thinspace s}}

\def\ergps{\hbox{$\erg\s^{-1}\,$}}

\def\kmps{\hbox{$\km\s^{-1}\,$}}
\def\kmpspMpc{\hbox{$\km\s^{-1}\Mpc^{-1}\,$}}

\def\pcmsq{\hbox{$\cm^{-2}\,$}}

\shorttitle{\sc{Silicate absorption in Compton-thick AGN}}
\shortauthors{{\sc A.D.~Goulding et al.}}

\begin{document}


\title{Deep silicate absorption features in Compton-thick AGN \\ predominantly arise due to dust in the host galaxy}


\author{A. D. Goulding\altaffilmark{1},
  D. M. Alexander\altaffilmark{2}, F. E. Bauer\altaffilmark{3}, W. R. Forman\altaffilmark{1},
  R. C. Hickox\altaffilmark{4},  \\ C. Jones\altaffilmark{1},
  J. R. Mullaney\altaffilmark{2,5}, M. Trichas\altaffilmark{1}}
\email{agoulding@cfa.harvard.edu}

\altaffiltext{1}{Harvard-Smithsonian Center for Astrophysics, 60
  Garden St., Cambridge, MA 02138, USA}
\altaffiltext{2}{Department of Physics, University of Durham, South
  Road, Durham DH1 3LE, UK}
\altaffiltext{3}{Departamento de Astronomia y Astrofisica, Pontificia
  Universidad Catolica de Chile, Casilla 306, Santiago 22, Chile}
\altaffiltext{4}{Department of Physics and Astronomy, Dartmouth
  College, Hanover, NH 03755, USA}
\altaffiltext{5}{Laboratoire AIM, CEA/DSM-CNRS-Universite Paris
  Diderot, Irfu/Service Astrophysique, CEA-Saclay, Orme des Merisiers,
  91191 Gif-sur-Yvette Cedex, France}

\begin{abstract}
  We explore the origin of mid-infrared (mid-IR) dust extinction in
  all 20 nearby ($z < 0.05$) {\it bona-fide} Compton-thick ($N_{\rm H}
  > 1.5 \times 10^{24} \pcmsq$) AGN with hard energy ($E > 10$~keV)
  X-ray spectral measurements. We accurately measure the silicate
  absorption features at $\lambda \sim 9.7\mu m$ in archival
  low-resolution ($R \sim 57$--127) {\it Spitzer} Infrared
  Spectrograph (IRS) spectroscopy, and show that only a minority
  ($\approx 45$\%) of nearby Compton-thick AGN have strong
  Si-absorption features ($S_{\rm 9.7} = ln(f_{\rm int}/f_{\rm obs})
  \gtrsim 0.5$) which would indicate significant dust attenuation. The
  majority ($\approx 60$\%) are star-formation dominated
  (AGN:SB$<$0.5) at mid-IR wavelengths and lack the spectral
  signatures of AGN activity at optical wavelengths, most likely
  because the AGN emission-lines are optically-extinguished. Those
  Compton-thick AGN hosted in low-inclination angle galaxies exhibit a
  narrow-range in Si-absorption ($S_{\rm 9.7} \sim 0$--0.3), which is
  consistent with that predicted by clumpy-torus models. However, on
  the basis of the IR spectra and additional lines of evidence, we
  conclude that the dominant contribution to the observed mid-IR dust
  extinction is dust located in the host galaxy (i.e., due to
  disturbed morphologies; dust-lanes; galaxy inclination angles) and
  not necessarily a compact obscuring torus surrounding the central
  engine.
\end{abstract}


\section{Introduction} \label{sec:intro}

The unified model of active galactic nuclei (AGN; e.g.,
\citealt{antonucci93}) is crucial to our understanding of the growth
and evolution of massive black holes and their host galaxies. A key
issue in AGN physics is the nature of the obscuring medium surrounding
the central engine.

There are three competing models for the specific structure, geometry
and composition of the obscuring material surrounding the central
supermassive black hole (SMBH): (1) a uniform (smooth/continuous),
heavily-obscuring, sub-parsec scale torus \citep{pier92,pier93}; (2)
an extended torus yielding moderate obscuration
\citep{granato94,efstathiou95,granato97}; and (3) clumpy `torus' of
many individual optically thick clouds
\citep{nenkova02,nenkova08}. Each of these theoretical models make
clear predictions for the observed torus properties. The
graphite/silicate dust contained within the torus is predicted to
extinguish AGN emission at UV/optical wavelengths, and should be
co-spatial with cool neutral gas which readily absorbs the X-ray
emission from the AGN (e.g., \citealt{mushotzky93}). However, this
dust/gas-rich torus is also predicted to (isotropically) re-emit in
the mid-infrared (mid-IR; $\lambda \sim 5$--$50\mu m$) with a mid-IR
spectral energy distribution characterized by a powerlaw-like AGN
continuum, superposed with silicate absorption/emission features at
9.7$\mu m$ and 18$\mu m$. The strength of these dust features are
expected to be dependent on the specific geometry and optical depth of
the torus (e.g., \citealt{fritz06,schartmann08}). To distinguish
between these torus models, we require well-constrained results from
sensitive and high-quality observations.


In nearby optical and radio-selected AGN, a weak correlation between
gas column density ($N_{\rm H}$) and silicate (Si) absorption strength
has been observed from {\it Spitzer} mid-IR spectroscopy (e.g.,
\citealt{shi06,wu09}). In turn, this suggests that selection of
sources with strong Si-absorption is a good method to find the most
heavily obscured AGN, i.e., Compton-thick sources with $N_H > 1.5
\times 10^{24} \pcmsq$ (e.g., \citealt{georgantopoulos11b}). To
first-order, these results would appear to agree with those predicted
by simple uniform torus models. However, there is growing evidence
that not all of the dust-extinction can be attributed to an obscuring
central torus.

Using ground-based high-spatial resolution photometry, \cite{Gandhi09}
show that Compton-thick AGN do not require significant corrections for
dust extinction to their AGN-produced mid-IR continuum, suggesting
that Si-absorption features in Compton-thick AGN are not being
produced in their nuclear regions (see also
\citealt{hoenig10,asmus11}). Furthermore, those AGN with strong
Si-absorption features are often found to be hosted in highly inclined
and/or merging galaxies (e.g., \citealt{deo07,deo09}), providing
first-order evidence that significant dust attenuation can occur
within the AGN host-galaxy (e.g., \citealt{alonso11} and references
there-in). Moreover, dust in the host-galaxy appears to extinguish the
optical emission-line signatures in $\approx 25$--50\% of nearby AGN
(e.g., \citealt{GA09}), and in a minority of cases, even the mid-IR
AGN emission-lines (e.g., NGC~4945; \citealt{goulding10}). This
apparent ambiguity among recent observations and theory raises two
fundamental questions: (1) is Si-absorption a common feature amongst
Compton-thick AGN, as predicted by a unified AGN model? and (2) does
Si-absorption predominantly arise from dust in a central torus or from
within the AGN host-galaxy?

To address these questions we explore the origins of apparent high
optical-depth dust in the sample of all 20 nearby ($z \lesssim 0.05$)
Compton-thick AGN which are unambiguously identified with high-energy
X-ray spectral observations. We use {\it Spitzer}-IRS observations to
determine the average mid-IR spectral energy distribution of
Compton-thick AGN, and establish whether silicate absorption is a
signature common to the most heavily-obscured AGN. In \S2 we describe
the sample of Compton-thick AGN, outline the data reduction methods of
the {\it Spitzer}-IRS observations and present the mid-IR spectra for
the Compton-thick AGN sample. In \S3 we compare the absorbing gas
column and the apparent dust-extinction levels towards the central AGN
and present the physical implications of these in light of a clumpy
torus model. Finally, in \S4 we review our results, finding that
the mid-IR Si-absorption features observed in Compton-thick AGN are
produced primarily due to dust in the AGN host-galaxy and not within a
central torus.

\section{Compton-thick AGN Sample \& Data Analyses} \label{sec:sample}

Compton-thick AGN ($N_{\rm H} > 1.5 \times 10^{24} \pcmsq $) are the
most heavily obscured class of AGN; by their very nature, they are
extremely difficult to detect and remain hidden in most X-ray surveys
(e.g.,
\citealt{Norman02,dma08,dma11,comastri11,feruglio11,gilli11,luo11}). Conclusive
identifications of Compton-thick AGN are made through spectroscopic
X-ray observations performed at $E > 10$~keV where the relatively
unabsorbed high-energy emission can be directly detected. The current
sensitivities of $E > 10$~keV observatories (e.g., {\it Beppo-SAX},
{\it Swift}, {\it Suzaku}, {\it INTEGRAL}) are substantially limited
by high backgrounds, relatively small effective areas and low spatial
resolution. To date, only 20 {\it bona-fide} Compton-thick AGN have
been unambiguously identified in the Universe at $E > 10$~keV (for a
review see \citealt{dellaceca08} and references there-in;
\citealt{awaki09,braito09}). These 20 Compton-thick AGN are all nearby
($z \lesssim 0.05$) systems hosted in spiral (Hubble-type S0 or later)
or merging galaxies with intrinsic X-ray luminosities $L_{\rm
  X,intrinsic} \sim (0.06$--$200) \times 10^{42} \ergps$ ($E \sim
2$--10~keV), and hence span the wide range of AGN power observed in
local Seyfert systems.

\begin{table*}
\tiny
\begin{center}
\setlength{\tabcolsep}{0.5mm}
\caption{The Compton-thick AGN sample\label{tbl:sample}}
\begin{tabular}{lrrrrccccccccrcccc}
\tableline\tableline
\multicolumn{1}{c}{Common} &
\multicolumn{1}{c}{$\alpha_{\rm J2000}$} &
\multicolumn{1}{c}{$\delta_{\rm J2000}$} &
\multicolumn{1}{c}{$z$} &
\multicolumn{1}{c}{$D_L$} &
\multicolumn{1}{c}{Gal.} &
\multicolumn{1}{c}{$b/a$} &
\multicolumn{1}{c}{D-L/} &
\multicolumn{1}{c}{Opt.} &
\multicolumn{1}{c}{$L_{\rm X,corr}$} &
\multicolumn{1}{c}{$N_{\rm H}$} &
\multicolumn{1}{c}{X-ray} &
\multicolumn{1}{c}{IRS} &
\multicolumn{1}{c}{AOR} &
\multicolumn{1}{c}{Obs.} &
\multicolumn{1}{c}{$S_{\rm 9.7}$} &
\multicolumn{1}{c}{$S_{\rm 9.7}$} &
\multicolumn{1}{c}{Mid-IR} \\

\multicolumn{1}{c}{Name} &
\multicolumn{1}{c}{} &
\multicolumn{1}{c}{} &
\multicolumn{1}{c}{} &
\multicolumn{1}{c}{(Mpc)} &
\multicolumn{1}{c}{Morph.} &
\multicolumn{1}{c}{} &
\multicolumn{1}{c}{Pec.} &
\multicolumn{1}{c}{Class} &
\multicolumn{1}{c}{(erg s$^{-1}$} &
\multicolumn{1}{c}{($\times 10^{24}$} &
\multicolumn{1}{c}{Ref.} &
\multicolumn{1}{c}{Type} &
\multicolumn{1}{c}{\#} &
\multicolumn{1}{c}{Date} &
\multicolumn{1}{c}{DecompIR} &
\multicolumn{1}{c}{\sc{PAHFIT}} &
\multicolumn{1}{c}{AGN:SB} \\

\multicolumn{1}{c}{} &
\multicolumn{1}{c}{} &
\multicolumn{1}{c}{} &
\multicolumn{1}{c}{} &
\multicolumn{1}{c}{} &
\multicolumn{1}{c}{} &
\multicolumn{1}{c}{} &
\multicolumn{1}{c}{} &
\multicolumn{1}{c}{} &
\multicolumn{1}{c}{cm$^{-2}$)} &
\multicolumn{1}{c}{$\pcmsq$)} &
\multicolumn{1}{c}{} &
\multicolumn{1}{c}{} &
\multicolumn{1}{c}{} &
\multicolumn{1}{c}{} &
\multicolumn{1}{c}{} &
\multicolumn{1}{c}{} &
\multicolumn{1}{c}{ratio} \\

\multicolumn{1}{c}{(1)} &
\multicolumn{1}{c}{(2)} &
\multicolumn{1}{c}{(2)} &
\multicolumn{1}{c}{(3)} &
\multicolumn{1}{c}{(4)} &
\multicolumn{1}{c}{(5)} &
\multicolumn{1}{c}{(6)} &
\multicolumn{1}{c}{(7)} &
\multicolumn{1}{c}{(8)} &
\multicolumn{1}{c}{(9)} &
\multicolumn{1}{c}{(10)} &
\multicolumn{1}{c}{(11)} &
\multicolumn{1}{c}{(12)} &
\multicolumn{1}{c}{(13)} &
\multicolumn{1}{c}{(14)} &
\multicolumn{1}{c}{(15)} &
\multicolumn{1}{c}{(16)} &
\multicolumn{1}{c}{(17)} \\

\tableline
NGC424            &  01h11m27.6s & -38d05m00s & 0.0125 &   50.8  & SBa  & 0.45  & - & Sy1/2   &  $42.63$ & $3.5^{+1.8}_{-1.4}$   & 1,2  & M    & 12444160     & 12-08-04  & $<0.00$  & 0.00 & $> 0.9$\\
NGC1068           &  02h42m40.7s & -00d00m47s & 0.0116 &   13.7  & SAb  & 0.85  & - & Sy2   & $>41.71$ & $>10.0$             & 3,4,5,6,7  & M    & 12461568     & 01-12-05  & 0.08  & 0.12 & $>0.9$ \\
E005-G004       &  06h05m41.6s & -86d37m55s & 0.0062 &   26.7  & Sb   & 0.21  & - & HII   &  $41.92$ & $1.6^{+0.5}_{-0.4}$   & 8  & S  & 18947328     & 10-25-06  & 0.61  & 0.47 & 0.60 \\
Mrk3              &  06h15m36.4s & +71d02m15s & 0.0134 &   60.6  & S0   & 0.89  & - & Sy2   &  $43.51$ & $1.3^{+0.2}_{-0.2}$   & 9,10,11  & S  & 3753472      & 03-04-04  & 0.33  & 0.15 & $>0.9$ \\
NGC2273           &  06h50m08.7s & +60d50m44s & 0.0038 &   26.5  & SBa  & 0.78  & - & Sy2   &  $42.23$ & $\sim 1.5$          & 12  & S  & 4851712      & 10-03-04  & 0.24  & 0.20 & 0.49 \\
NGC3079           &  10h01m57.8s & +55d40m47s & 0.0061 &   16.2  & SBc  & 0.18  & - & L/Sy2 &  $42.10$ & $10.0^{+10.0}_{-5.3}$ & 13  & S  & 3755520      & 04-19-04  & 0.65  & 0.58 & 0.32 \\
NGC3281           &  10h31m52.1s & -34d51m13s & 0.0038 &   49.6  & SAab & 0.52  & Y & Sy2   &  $43.17$ & $2.0^{+0.2}_{-0.1}$   & 4,10,14  & S  & 4852224      & 05-23-05  & 1.41  & 1.28 & 0.87 \\
NGC3393           &  10h48m23.5s & -25d09m43s & 0.0115 &   56.2  & SB0  & 0.91  & - & Sy2   & $>42.85$ & $>10.0$             & 7,15  & S  & 4852480      & 05-22-05  & 0.13  & 0.14 & 0.77 \\
Arp299            &  11h28m30.4s & +58d34m10s & 0.0110 &   44.8  & Irr  & -  & Y & HII   &  $42.84$ & $2.5^{+1.4}_{-0.6}$   & 16,17  & M    & 3840256      & 04-15-04  & 0.90  & 0.92 & 0.63 \\   
Mrk231            &  12h56m14.2s & +56d52m25s & 0.0420 &  186.0  & SAc  & 0.77  & - & QSO   &  $43.70$ & $\sim 2.0$          & 18  & S  & 4978688      & 04-14-04  & 0.18  & 0.13 & 0.77 \\
NGC4939           &  13h04m14.4s & -10d20m22s & 0.0104 &   44.8  & SAbc & 0.51  & - & Sy2   & $>42.63$ & $>10.0$             & 15  & S  & 4853760      & 01-06-04  & 0.00  & 0.00 & $>0.9$ \\
NGC4945           &  13h05m27.5s & -49d28m05s & 0.0019 &    3.9  & Scd  & 0.19  & Y & HII   &  $42.30$ & $2.2^{+0.3}_{-0.4}$   & 4,5,6,19  & S  & 8768928     & 03-01-04  & 1.40  & 0.74 & $<0.2$ \\
NGC5194           &  13h29m52.7s & +47d11m42s & 0.0015 &    8.6  & SAbc & 0.99  & - & L/Sy2 &  $40.76$ & $5.6^{+4.0}_{-1.6}$   & 20  & M    & 9480192      & 05-12-04  & 0.00  & 0.04 & 0.43 \\
Circinus          &  14h13m09.9s & -65d20m20s & 0.0014 &    3.7  & SAb  & 0.43  & Y & Sy2   &  $41.90$ & $4.3^{+0.4}_{-0.7}$   & 11,21,22  & S  & 9074176      & 03-01-04  & 1.28  & 1.25 & 0.88 \\
NGC5728           &  14h42m23.9s & -17d15m11s & 0.0094 &   40.9  & SABa & 0.58  & - & Sy2   &  $43.04$ & $2.1^{+0.2}_{-0.2}$   & 10,23  & S  & 18945536     & 08-05-07  & 0.14  & 0.12 & 0.52 \\
E138-G001       &  16h51m20.1s & -59d14m05s & 0.0091 &   38.6  & S0   & 0.99  & - & Sy2   &  $42.52$ & $\sim 1.5$          & 2,6  & S  & 17643264     & 04-30-07  & 0.00  & 0.00 & $> 0.9$\\
NGC6240           &  16h52m58.9s & +02d24m03s & 0.0243 &  105.0  & Irr  & -  & Y & L     &  $44.26$ & $2.2^{+0.4}_{-0.3}$   & 11,24  & S  & 4985600      & 03-04-04  & 0.82  & 0.62 & 0.44 \\
IRAS19254-7245    &  19h31m21.4s & -72d39m18s & 0.0620 &  266.0  & Irr  & -  & Y & QSO & $44.50$ & $3.1^{+1.2}_{-0.4}$ & 27 & S & 12256512 & 05-30-05 & 1.07 & 0.58 & 0.64 \\
NGC7582           &  23h18m23.5s & -42d22m14s & 0.0053 &   22.7  & SBab & 0.42  & Y & Sy2   &  $42.61$ & $1.6^{+0.9}_{-0.5}$   & 10,25  & M    & 12445184     & 05-25-05  & 0.76  & 0.79 & 0.37 \\
NGC7674           &  23h27m56.7s & +08d46m44s & 0.0289 &  127.0  & SAbc & 0.91  & - & Sy2   & $>43.34$ & $>10.0$             & 26  & M    & 12468736     & 12-11-04  & 0.10  & 0.00 & $>0.9$\\
\tableline
\end{tabular}
\end{center}
\begin{spacing}{0.5}
{\bf Notes:-} $^{1}$Common galaxy name;
$^{2}$J2000 positional co-ordinates from the NASA Extragalactic Database (NED);
$^{3}$Spectroscopic redshift;
$^{4}$Luminosity distance in Megaparsecs assuming $H_0 = 71 \kmpspMpc$ and $\Omega_{\Lambda} = 0.70$ and corrected for non-cosmological flows;
$^{5-6}$Galaxy morphology and major/minor axis ratio taken from the Third Reference Catalog of Bright Galaxies \citep{rc3};
$^{7}$Y -- denotes dust-lanes or peculiar morphology are evident in the optical image of the source;
$^{8}$Optical classification in NED;
$^{9}$Logarithm of intrinsic (2--10 keV) X-ray luminosity in units of $\ergps$ corrected for Compton-thick absorption for those sources with $N_{\rm H} < 10^{25} \pcmsq$;
$^{10}$Neutral hydrogen column density derived from hard X-ray spectroscopy in units of $10^{24} \pcmsq$;
$^{11}$Reference for X-ray data (see below);
$^{12}$Type of {\it Spitzer}-IRS observation (M: Mapping; S: Staring);
$^{13}${\it Spitzer} observation record indicator; 
$^{14}$Date of {\it Spitzer}-IRS observation;
$^{15}$Depth of Si-absorption at $\lambda \sim 9.7 \mu m$
derived from spectral fits to {\it Spitzer}-IRS data using the formalism of
Mullaney et al. (2011);
$^{16}$Depth of Si-absorption at $\lambda \sim 9.7 \mu m$
derived from spectral fits to Spitzer-IRS data using PAHFIT;
$^{17}$AGN--starburst ratio derived from fitting IRS spectrum with {\tt DecompIR} (AGN-dominated system $= 1$).\\
{\bf References:-}
(1) \cite{iwasawa01}; 
(2) \cite{collinge00}; 
(3) \cite{matt97}; 
(4) \cite{sazonov07}; 
(5) \cite{beckmann06};
(6) \cite{bassani07};
(7) \cite{levenson06}; 
(8) \cite{ueda07}; 
(9) \cite{cappi99};
(10)\cite{markwardt05};	
(11) \cite{bassani99}; 
(12) \cite{awaki09};
(13) \cite{iyomoto01};
(14) \cite{vignali02};
(15) \cite{maiolino98};
(16) \cite{dellaceca02};
(17) \cite{ballo04};
(18) \cite{braito04};
(19) \cite{guainazzi00}; 
(20) \cite{fuzakawa01};
(21) \cite{matt99};
(22) \cite{iwasawa97};
(23) \cite{comastri07};
(24) \cite{vignati99};
(25) \cite{turner00};
(26) \cite{malaguti98};
(27) \cite{braito09}
\end{spacing}
\normalsize
\end{table*}

\subsection{{\it Spitzer}-IRS Data Reduction}

All 20 bona-fide Compton-thick AGN identified to date have archival
low-resolution {\it Spitzer}-IRS spectroscopy. Specifically, these 20
AGN have been observed with the low-resolution modules (short-low [SL;
5.2--$14.5 \mu m$] and long-low [LL; 14.0--$38.0 \mu m$]; $R \approx
57$--127) in either staring or mapping mode as part of multiple
programs, and hence these data form a heterogeneous, but still
complete sample. Our sample of nearby Compton-thick AGN and their
basic properties are given in Table 1. The mid-IR data for many of our
AGN sample have been analyzed using different reduction techniques in
previous papers (e.g.,
\citealt{deo07,hao07,wu09,mullaney10,mullaney11}; Sazonov et al. 2012,
submitted). However, in order to ensure self-consistency for the
measurement of the spectral absorption/emission features which are
integral to the analyses presented here, we have re-extracted each of
the IRS observations using our own custom {\sc idl} reduction routine
(see \citealt{goulding_phd10,goulding11,mullaney11}).

Briefly, the two-dimensional Basic Calibrated Data (BCDs) images,
produced by the S18.18.0 {\it Spitzer} Science Center (SSC) pipeline,
were retrieved and rigorously cleaned of rogue `hot' pixels using our
customized version of {\sc irsclean}. Next, individual rows were fit
as a function of time to remove latent charge which exists on the
detector after bright and/or long observations. The IRS Staring
observations were averaged in the different nod positions, which were
then used to perform alternate background subtractions of the source
in each nod position and final spectra were extracted using the {\it
  Spitzer}-IRS Custom Extraction ({\sc spice}) software provided by
the SSC. Cleaned and processed IRS Mapping observations were input to
the SSC package {\sc cubism} to build the spectral data cubes and
extract the 1-D spectroscopy. Individual IRS orders in the 1-D spectra
were clipped of noise (see the {\it Spitzer}-IRS handbook for further
information) and stitched together by fitting low-order polynomials to
produce the final continuous spectra for each source. In the
upper-panel of Fig.~\ref{fig:irs_spec}, we show the resulting
low-resolution {\it Spitzer}-IRS spectroscopy for our complete sample
of nearby Compton-thick AGN.

\subsection{{\it Spitzer}-IRS Spectral Decomposition \& Silicate
  Absorption Measurement}

On the basis of uniform dust torus models produced from radiative
transfer theory (e.g., \citealt{dullemond05}; Schartmann et
al. 2005; Fritz et al. 2006), the mid-IR spectral energy distributions
of intrinsically heavily-absorbed Type-2 AGN are expected to be
dominated by an AGN-produced powerlaw and most, if not all, are
expected to exhibit significant Si-absorption features at $\lambda
\sim 9.7 \mu m$. Throughout this manuscript, we define $S_{\rm 9.7}$
as the depth of the Si-absorption feature at $\lambda \sim 9.7 \mu m$
where $S_{\rm 9.7} = ln(f_{\rm 9.7,intrinsic}/f_{\rm
  9.7,observed})$. Following many previous studies, we use $S_{\rm
  9.7}$ as a good proxy for the apparent optical depth and
dust-extinction in our sample (e.g.,
\citealt{spoon07,shi06,levenson07,deo07,deo09,georgantopoulos11b}). Though
we note that the relation between $S_{\rm 9.7}$ and true optical depth
is most-likely non-linear and somewhat model and orientation dependent
(see \citealt{levenson07,nenkova08,schartmann08}).

\begin{figure}[ht]
\centering
\includegraphics[width=0.98\linewidth]{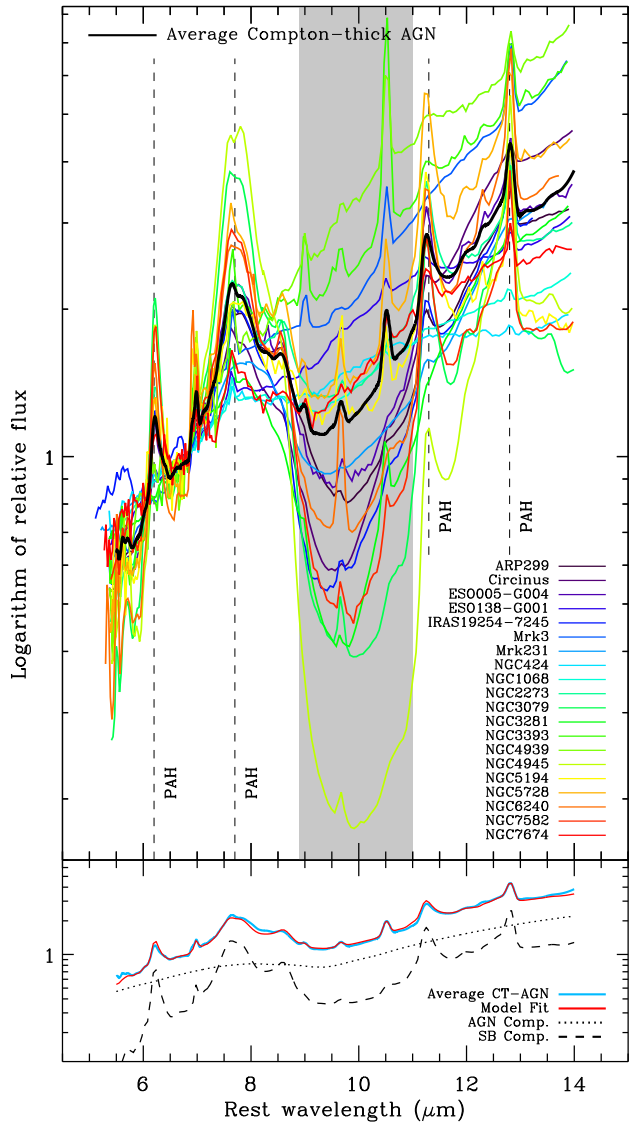}
\vspace{-0.5cm}
\caption{{\bf Upper Panel:} Low resolution ($R \sim 57$--127) {\it
    Spitzer}-IRS spectroscopy of all 20 bona-fide Compton-thick AGN in
  the nearby Universe with high-energy ($E> 10$~keV) spectroscopic
  constraints. The gray shaded region represents the wavelength
  boundaries for the dominant Si-absorption feature ($\lambda \sim 9.7
  \mu m$) in the short-low (SL) {\it Spitzer}-IRS pass-band. The
  mid-IR spectral energy distributions of Compton-thick AGN are
  clearly very diverse with spectra dominated by a combination of an
  AGN-produced powerlaw; strong starburst-produced polycyclic aromatic
  hydrocarbon features at $\lambda \sim 6.3$, 7.7, 11.3 and $12.8\mu
  m$; and in a {\it minority} (9/20) of cases, strong ($S_{\rm 9.7} >
  0.3$) Si-absorption features. Additionally, we show the average
  mid-IR spectrum of the sample of Compton-thick AGN (thick black
  line). {\bf Lower panel:} Spectral decomposition of the average
  Compton-thick AGN spectrum (blue solid curve) The best-fit
  extinction-convolved AGN power-law and starburst template are shown
  with dotted and dashed curves, respectively. The total best-fit
  spectrum (i.e., power-law $+$ starburst $+$ emission-lines) is shown
  with a solid red curve. The best-fitting constraints for various AGN
  classifications are given in Table 2.}
\label{fig:irs_spec}
\end{figure}

Previous investigations have often employed simple extrapolation
methods to estimate the depth of the silicate feature in mid-IR
spectroscopy (e.g.,
\citealt{spoon07,levenson07,georgantopoulos11b}). This involves
measurement of the observed continuum blue-ward and red-ward of the
silicate feature, extrapolating a powerlaw between the continuum
points and comparing this to the observed flux at $\lambda \sim 9.7
\mu m$. In principle, this extrapolation method is sufficient for
AGN-dominated spectra. However, as shown in Fig.~\ref{fig:irs_spec},
the majority of the mid-IR spectra for Compton-thick AGN contain
strong polycyclic-aromatic hydrocarbon (PAH) features (at 6.3, 7.7,
11.3 and $12.3 \mu m$) indicative of substantial circumnuclear
star-formation activity. These PAH features provide significant
contributions to the mid-IR spectra blue-ward of the Si-absorption
feature, having the effect of artificially increasing the continuum
strength at $\lambda \sim 6$--$8 \mu m$, and hence, over-estimating
the intrinsic AGN continuum flux at $\lambda \sim 9.7 \mu m$. In a
pure synthetic star-forming template with {\it no} evidence for
extinction, a substantial Si-absorption feature ($S_{\rm 9.7} \sim
0.7$) would be inferred using this basic method. Therefore, we choose
to use a spectral decomposition method to remove the host-galaxy
emission, and measure the depth of the Si-absorption feature only
within the AGN component. Furthermore, to constrain any systematic
uncertainties derived from using any one spectral decomposition method
we use two independent spectral decomposition procedures.

First, we use the {\sc idl} software package {\tt DecompIR} (Mullaney
et al. 2011), which performs a Chi-squared minimization of the
observed spectrum to a combination of starburst templates (i.e., a
host-galaxy component) and an absorbed power-law (i.e., an AGN
component). Here, we employ a simple screen extinction curve
\citep{draine03} to account for absorption of the AGN component; thus,
we do not attempt to constrain the physical region where dust
extinction is occurring. Following Goulding et al. (2011), we allow
for a range of star-forming templates, including that of the
archetypal nearby starburst, M82; a combined \cite{brandl06} starburst
template;\footnote{We do not include NGC~660, NGC~1365, NGC~3628 and
  NGC~4945 as part of the Brandl et al. (2006) template as these
  sources are known to harbor AGN.} and a range of compact nuclear and
extra-nuclear theoretical starburst templates
\citep{siebenmorgen07}. From the fitted AGN component, we predict the
intrinsic power of the AGN at $\lambda \sim 9.7 \mu m$, and compare
this to the observed luminosity to establish an estimate for $S_{\rm
  9.7}$ (column 15 of Table 1). Second, we employ the widely used
spectral-fitting package, {\tt PAHFIT} (\citealt{smith07}) to fit a
range of heated dust continua combined with PAH and emission line
features to derive $S_{\rm 9.7}$ (column 16 of Table 1). Following,
\cite{georgantopoulos11b}, we change the default dust continuum
temperatures in {\tt PAHFIT} and additionally allow hotter dust
temperatures (400--1400K) to be fit to the spectra, these are expected
to arise from an accretion heated AGN torus. In general, we find very
good agreement for $S_{\rm 9.7}$ derived using {\tt DecompIR} and {\tt
  PAHFIT} (see columns 15 and 16 of Table 1). In \S\ref{sec:extinct},
we compare these two measures of the Si-absorption feature to gas
column density in Compton-thick AGN.

\subsection{The Average mid-IR Spectra of Compton-thick AGN} \label{sec:avg_spec}

In Figure~\ref{fig:irs_spec} we show that Compton-thick AGN at $z \sim
0$ are characterized by a variety of spectral shapes with
AGN--starburst ratios of $\sim 0.1$--1. Of the 20 AGN in our sample,
only four appear to be dominated by a featureless power-law continuum,
while eleven show strong PAH features indicative of circumnuclear
starburst activity which, for seven ($\approx 35$\%) of the sources,
even dominates the bolometric output of the galaxies at mid-IR
wavelengths (AGN:SB $\lesssim 0.5$; see Column 17 of Table 1). By
contrast, Sazonov et al. (2012, submitted) find that only $\lesssim
20$\% of all hard X-ray selected Compton-thin ($N_{\rm H} \sim
(0.05$--$100) \times 10^{22} \pcmsq$) Type-2 AGN detected by {\it
  INTEGRAL} are dominated by star-formation at mid-IR wavelengths. It
is therefore possible that enhanced star-formation activity may be a
characteristic specific to the most heavily absorbed ($N_{\rm H} >
10^{24} \pcmsq$) AGN, and not merely a hard X-ray selection effect, at
least in the nearby Universe.

Through spectral stacking, we find that Compton-thick AGN, on average,
show strong star-formation activity. We stacked all 20 mid-IR spectra
to produce a mean mid-IR spectral energy distribution for
Compton-thick AGN. We chose to normalize the rest-frame mid-IR spectra
for each of the Compton-thick AGN at $\lambda \sim 6.5$--$7 \mu m$
(i.e., a wavelength range with a relatively featureless continuum). In
the lower panel of Fig.~\ref{fig:irs_spec}, we show the spectral
decomposition of the average spectrum modeled using {\tt
  DecompIR}. Our modified \cite{brandl06} template provides the best
$\chi^2$-fit `starburst' to the mean Compton-thick AGN spectrum, and
contributes $\approx 40$\% of the overall luminosity to Compton-thick
AGN at mid-IR wavelengths.

As we show in Table~1 and Fig.~\ref{fig:irs_spec}, the detection of
Si-absorption features in nearby Compton-thick AGN is far from
ubiquitous. The Compton-thick AGN cover a wide range in Si-absorption
depth, $S_{\rm 9.7} \sim 0$--1.4. While the mid-IR spectra for eight
of the Compton-thick AGN do appear to be moderately extinguished, we
also find that the spectral decomposition (using {\tt DecompIR}) of
NGC~424 shows strong evidence for Si-{\it emission} ($S_{\rm 9.7} <
0.0$); based on unified models, Si-emission features are expected only
in unobscured Type-1 AGN, irrespective of specific torus
geometry/composition. This result is also partially confirmed by {\tt
  PAHFIT} which produces a relatively poor fit ($\chi^2_{n-1} \sim
26.1$) for NGC~424 and by design does not allow apparent optical
depths of $S_{\rm 9.7} < 0.0$. Explanations for Si-emission features
in Type-2 AGN (e.g., NGC 2110, Mason et al. 2009; SST1721+6012,
\citealt{nikutta09}) are that the emission is either (i) arising from
the innermost region of the AGN narrow-line region, above the
scale-height of the torus \citep{mason09}; or (ii) from an unobscured
inner region of a low-mass clumpy torus (e.g., based on the models of
Schartmann et al. 2008).

From the AGN powerlaw component of the stacked Compton-thick AGN
spectrum, we directly measure the depth of the Si-absorption feature
at $\lambda \sim 9.7 \mu m$. We find that, on average, the re-radiated
AGN emission of a Compton-thick source experiences an apparent
attenuation of only $S_{\rm 9.7} \sim 0.36 \pm 0.04$, equivalent to an
optical extinction of $A_{\rm V} \approx 3.9$--4.9
mags.\footnote{Throughout we assume a standard optical-IR extinction
  curve \citep{draine07} in the direction of the Milky-way galactic
  center such that $A_{9.7\mu m} / A_{\rm V} \simeq 0.075$.} Optical
depths such as these are readily observed in nearby galactic
star-forming regions (e.g., Draine 2003). In \S~\ref{sec:host_gal}, we
further explore whether this measured dust extinction in Compton-thick
AGN can be attributed only to that arising in the host-galaxy and not
necessarily from an obscuring torus.

\section{Gas Absorption and Dust Extinction in Compton-thick AGN} \label{sec:extinct}

\begin{figure*}[ht]
\centering
\includegraphics[width=0.96\textwidth]{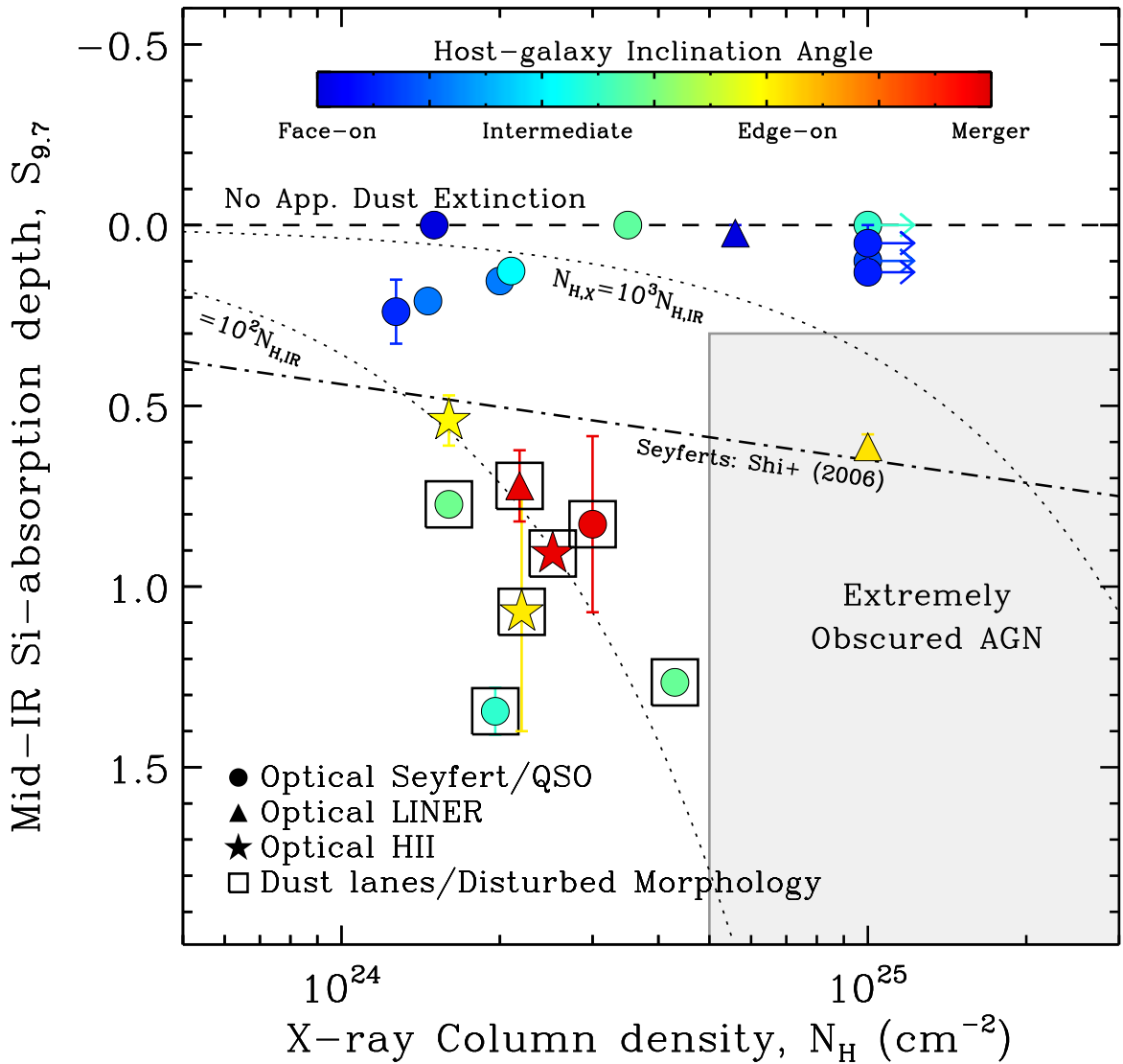}
\vspace{-0.5cm}
\caption{Si-absorption depth at $9.7\mu m$ ($S_{\rm 9.7}$; a good
  proxy for apparent dust extinction) observed in {\it Spitzer}-IRS
  mid-IR spectroscopy versus hard X-ray derived gas column density
  ($N_{\rm H}$) in units of $\pcmsq$ for all bona-fide Compton-thick
  AGN in the nearby Universe. Error bars represent the range in
  $S_{\rm 9.7}$ measured using {\tt DecompIR} and {\tt
    PAHFIT}. Constant gas--dust ratios are shown (dotted lines) for
  ratios of $N_{\rm H,X}/N_{\rm H,IR} = 100$ and 1000 assuming $N_{\rm
    H} / A_{\rm V} = 1.93 \times 10^{21} \pcmsq$ and $A_{\rm 9.7} /
  A_{\rm V} \simeq 0.075$ (e.g., \citealt{draine07}). Inclination
  angles are shown with color-gradient and are derived from
  major--minor axis ratios (see Column 6 of Table 1), face-on systems
  ($b/a \sim 1$) are shown in blue, edge-on ($b/a \sim 0$) in yellow,
  colors in between are linearly scaled to intermediate inclination
  angles ($b/a \sim 0$--1), and merging systems (established from
  optical imaging and NED classifications) are shown in red. Optical
  spectral classifications (AGN; LINER; HII/star-forming) are shown
  with filled circles, triangles and stars,
  respectively. Additionally, sources with dust-lanes or disturbed
  optical morphologies are highlighted with open squares (see Column 7
  of Table 1). For comparison, we show the linear-fit presented in Shi
  et al. (2006) for a sample of Seyfert galaxies with measured $N_{\rm
    H}$.}
\label{fig:nhtau}
\end{figure*}

Unified AGN models postulate that obscuring gas and dust are both
located in the circumnuclear torus, so that large gas columns are
associated with heavy dust extinction and, for a uniform torus, deep
Si-absorption features. With this in mind, a number of recent studies
use the existence of high apparent optical depth ($S_{\rm 9.7} > 1$)
to identify Compton-thick AGN (e.g.,
\citealt{georgantopoulos11b,nardini11,dasyra11}). However, as shown in
Fig.~1, the large range in $S_{\rm 9.7}$ seen across our complete
Compton-thick AGN sample suggests that the dust producing the
Si-absorption features may not be co-spatial with the X-ray absorbing
gas.

\subsection{Compton-thick AGN with little apparent dust
  extinction} \label{sec:gas_vs_dust}

To establish if a physical link exists between the gas and dust in
Compton-thick AGN, as expected by a unified model, in this section, we
compare the depth of the Si-absorption feature and the gas column
density. In Fig.~\ref{fig:nhtau}, we plot the measured X-ray gas
column density versus mid-IR Si-absorption strength ($S_{\rm
  9.7}$). For each AGN, we use the average value for $S_{\rm 9.7}$
derived independently from {\tt DecompIR} and {\tt PAHFIT} in the
previous section. Error bars represent the spread in the values of
$S_{\rm 9.7}$ for a particular source. We find little or no observed
correlation between gas absorption and dust extinction in
Compton-thick AGN (Spearman's rank, $\rho \sim -0.30$, $P_{\rm null}
\sim 0.21$). At least eight of the sources exhibit little or no
evidence for dust extinction ($S_{\rm 9.7} \lesssim 0.1$), and
moreover, NGC~424 has silicate {\it emission} features ($S_{\rm 9.7} <
0.0$), in general, such emission features require a direct
line-of-sight to the inner AGN torus. By contrast, 9 of the 20 ($\sim
45 \pm 18$\%) Compton-thick AGN in our sample appear moderately
($S_{\rm 9.7} \sim 0.5$--1.0) to heavily ($S_{\rm 9.7} > 1.0$)
obscured at mid-IR wavelengths, as would be expected by unified AGN
models that invoke a smooth uniform dust torus. However, the fraction
of Compton-thick AGN with high apparent optical depth ($S_{\rm 9.7}
\gtrsim 0.5$) is similar to that measured for lower column density
($N_{\rm H} \sim (0.05$--$50) \times 10^{22} \pcmsq$) AGN detected by
{\it Swift}-BAT, $30 \pm 10$\% (derived from the spectroscopy
presented in \citealt{mullaney10}), suggesting that large $N_{\rm H}$
does not necessarily predicate large dust extinction. Indeed, at least
four ($\approx 10$\%) Compton-thin Swift-BAT AGN presented in
\cite{mullaney10} have $S_{\rm 9.7} \gtrsim 1$.

In Fig.~\ref{fig:nhtau} we additionally show curves of constant gas
column density predicted using $S_{\rm 9.7}$ assuming a standard
optical--IR Milky-way extinction curve \citep{draine07} and a
gas--dust ratio of $N_{\rm H}/A_{\rm V} = 1.93 \times 10^{21} \pcmsq
{\rm mag}^{-1}$ (\citealt{fitzpatrick85,draine03} for $R_V = A_V /
E(B-V) = 3$). We show that Compton-thick AGN are required to have
gas--dust ratios which are a factor $\approx 50$--10,000 greater than
observed galactically, which may suggest there is no single gas--dust
ratio in these sources (e.g.,\citealt{maiolino01}). These large
gas--dust ratios may be explained by additional absorption of X-ray
emission by large quantities of neutral gas residing within the
dust-sublimation radius ($0.1 < r < 10$~pc) of the obscuring torus,
so-called `partial covering'.

Recent X-ray monitoring campaigns of variable absorption state AGN
(e.g., Mrk766; NGC~1365; NGC~4151) have observed short-term
occultations of the X-ray source by high-velocity ($\sim 3000 \kmps$)
circumnuclear gas clouds residing at $\sim 0.1$~pc (e.g.,
\citealt{risaliti09,risaliti09c,wang10}). The motion and distance from
the central source of these gas clouds is consistent with that
expected for X-ray absorption arising from the AGN broad-line region
(BLR; e.g., \citealt{risaliti09c,risaliti11}). Within these relatively
small radii, SiO dust grains are preferentially destroyed due to the
high temperatures ($T > 1000$K). Hence, the gas-dust ratio of the BLR
may be far larger than that of the dusty-torus. However,
partial-covering may not explain the high gas--dust ratio observed in
all of the Compton-thick AGN included here. Recent, spatially-resolved
results of \cite{marinucci12b} suggest that the majority of the
Compton-thick absorption in NGC~4945 may be at far larger radii ($\sim
100$~pc) than the BLR. These variations in BLR and torus absorptions,
as well as inclination effects (e.g., \citealt{nenkova08}) may add
substantial scatter to an expected gas-dust ratio for nearby
AGN. Taken together, these results give rise to several interesting
scenarios for: (1) the spatial extent of the intervening gas and dust;
(2) the density gradient of the materials across the torus; and (3)
possible gas--dust separation boundaries within a torus. Given our
small sample and (by design) our limited range in $N_{\rm H}$, these
issues are beyond the scope of this investigation. We may still
robustly conclude that all Compton-thick AGN (i.e., large $N_{\rm H}$
sources) are {\it not} necessarily obscured by cool dust with high
apparent optical depth. However, the reverse of this statement may
still also be true, in that all systems observed to have $S_{\rm 9.7}
\gg 1$ may all contain Compton-thick AGN, but are currently missed due
to the lack of sufficiently sensitive instrumentation; although these
extreme $S_{\rm 9.7} \gg 1$ sources are only a limited subset ($<
20$\%) of the Compton-thick AGN population.

\subsection{Evidence for dust extinction arising in the host galaxy of
Compton-thick AGN} \label{sec:host_gal}

The apparent lack of a correlation between gas and dust attenuation in
Compton-thick AGN presented in the previous section suggests that the
gas and dust may not be co-spatial. Indeed, \cite{sturm05} suggested
an extranuclear origin for the Si-absorption features observed in both
low and high-luminosity AGN. This is supported by Gandhi et al. (2009)
showing in high spatial resolution ground-based photometry of nuclear
regions in nearby AGN, that the AGN-produced mid-IR continuum appears
almost isotropic and unaffected by Si-absorption. Furthermore, the
inclination angle of the host galaxy may have a significant obscuring
effect on the AGN emission (e.g.,
\citealt{malkan98,matt00,alonso11,lagos11}). In particular, those AGN
presenting strong Si-absorption features are often found to be hosted
in highly-inclined and/or merging galaxies
(\citealt{deo07,deo09}). For some host-galaxy types and inclinations,
the dust extinction may be so over-whelming that it has the ability to
completely obscure an AGN (e.g., at optical and even mid-IR
wavelengths; \citealt{GA09}). Indeed, the mid-IR AGN emission in the
late-type edge-on AGN, NGC 4945 is almost entirely obscured based on
{\it Spitzer}-IRS spectroscopy. Hence, in this section, we test for a
link between the observed mid-IR extinction and the properties of the
host galaxy in Compton-thick AGN.

\begin{table}
\footnotesize
\begin{center}
\setlength{\tabcolsep}{1.2mm}
\caption{Stacked properties of Compton-thick AGN\label{tbl:props}}
\begin{tabular}{lcc}
\tableline\tableline
\multicolumn{1}{c}{Sub-sample} &
\multicolumn{1}{c}{$S_{\rm 9.7}$} &
\multicolumn{1}{c}{AGN:SB} \\
\multicolumn{1}{c}{(1)} &
\multicolumn{1}{c}{(2)} &
\multicolumn{1}{c}{(3)} \\
\tableline

All sources & 0.36 & 0.62 \vspace{2mm}\\

Optical Seyfert* & 0.14 & 0.83 \\
Optical LINER/HII* & 0.55 & 0.43 \vspace{2mm}\\

Face-on/Intermediate Inclination & 0.19 & 0.75 \\
Edge-on Inclination or Merger & 0.96 & 0.44 \\
\tableline
\end{tabular}
\end{center}
\vspace{-0.2cm}
{\bf Notes:-}
$^{1}$Host galaxy properties of sub-sample of main Compton-thick AGN sample;
$^{2}$Average depth of Si-absorption feature ($S_{\rm 9.7}$) within stacked
mid-IR spectra of sub-sample, measured using {\tt DecompIR};
$^{3}$AGN--starburst ratio derived from decomposition of stacked
mid-IR spectra using {\tt DecompIR} (AGN-dominated $= 1$). $^*$Excluding QSOs.
\end{table}

In Fig.~\ref{fig:nhtau}, we highlight the optical classifications and
host-galaxy inclinations for the Compton-thick AGN
sample.\footnote{The host-galaxy inclination angles are derived from
  the major--minor axis ratio as defined in the Third Reference
  Catalog of Bright Galaxies \citep{rc3}.} As noted previously, each
of the 20 Compton-thick AGN are hosted in late-type (S0--Scd) or
irregular/merging galaxies. From stacked mid-IR spectra, we find that
the galaxies classified optically as Seyferts (i.e., using standard
optical emission-line diagnostic diagrams) have lower average apparent
dust extinction ($S_{\rm 9.7} \sim 0.14$) than the Compton-thick AGN
hosted in the optically-classified Low-Ionization Nuclear Emission
Line Regions (LINERs; i.e., sources with optical spectra
characteristic of LINERs) and HII galaxies ($S_{\rm 9.7} \sim
0.55$). This dichotomy between apparent dust extinction and galaxy
classification provides further evidence that the optical class may be
influenced by high-levels of intervening dust in the host
galaxy. Additionally, we show that Compton-thick AGN hosted in face-on
(low-inclination) galaxies have mid-IR spectra exhibiting little or no
apparent dust extinction ($S_{\rm 9.7} \sim 0.19$). By contrast,
Compton-thick AGN hosted in edge-on and/or merging systems have
significantly greater dust extinction ($S_{\rm 9.7} \sim 0.96$; see
Table 2) clearly suggesting that the Si-absorption features observed
at mid-IR wavelengths have an extra-nuclear origin.

\begin{figure}[ht]
\centering
\includegraphics[width=0.99\linewidth]{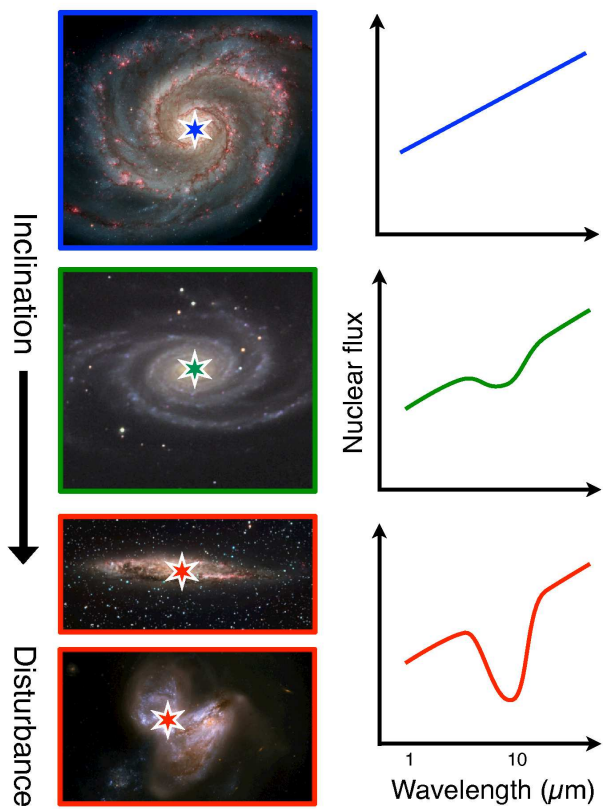}
\vspace{-0.5cm}
\caption{Schematic diagram illustrating the effect of host-galaxy
  inclination (and morphology disturbance) on the perceived dust
  extinction at mid-IR wavelengths. Based on the effects presented
  here, the strength of the Si-absorption feature at $\lambda \sim 9.7
  \mu m$ qualitatively appears to increase with increasing galaxy
  inclination, i.e., consistent with the view that the obscuring dust
  which dominates the mid-IR extinction features arises only from
  within the host galaxy. Image credits, from top: {\em NGC 5194}:
  NASA and the Hubble Heritage Team (STSci/AURA); {\em NGC 4939}: Adam
  Block/NOAO/AURA/NSF; {\em NGC 4945}: ESO; {\em Arp 299}: NASA and
  the Hubble Heritage Team (STSci/AURA)}
\label{fig:inclin}
\end{figure}

There are also three optically-classified Seyferts (NGC~3281;
NGC~4939; Circinus) within our Compton-thick sample which have larger
Si-absorption features ($S_{\rm 9.7} \sim 0.8$--1.5) than would
necessarily be expected given their moderate inclination angles ($b/a
\sim 0.5$). Visual inspection of optical images used to derive the
morphological classifications shows that these three Compton-thick AGN
have disturbed nuclear morphologies and/or dust-lanes aligned along
the galaxy nucleus. These additional extra-nuclear features may
provide further regions of cool-dust, and hence, larger Si-absorption
strengths. Taken together, these results suggest that the dominant
source of dust-extinction observed at mid-IR wavelengths is not
necessarily linked with an obscuring torus, but instead is arising
from extended kilo-parsec-scale regions which lie far outside the
expected dust-sublimation region of an accreting SMBH. Indeed, this
galactic scale origin for the cool dust, which is required to produce
deep Si-absorption features, is consistent with the wide range of
$S_{\rm 9.7}$ observed in nearby starburst-dominated Ultra-luminous IR
galaxies (e.g., \citealt{spoon07,desai07,farrah08}).

Despite our relatively small sample of sources, the 20 Compton-thick
AGN considered here cover a wide range of apparent dust extinction
values, $S_{\rm 9.7} \sim 0.0$--1.7. However, in Fig.~\ref{fig:nhtau}
we note a clear dearth of sources with $S_{\rm 9.7} > 0.3$ and $N_{\rm
  H} > 5 \times 10^{24} \pcmsq$. AGN in this regime are likely to be
heavily obscured at both mid-IR and optical wavelengths due, in part,
to large quantities of intervening dust within the host galaxy, as
well as being heavily obscured at X-ray energies due to large gas
columns within the torus. Such heavily-obscured AGN are still not
robustly identified in the nearby Universe. Observational expectations
are that the number of Compton-thick AGN with $N_{\rm H} \sim (1$--$5)
\times 10^{24} \pcmsq$ and $N_{\rm H} > 5 \times 10^{24} \pcmsq$ are
roughly constant (e.g., \citealt{risaliti99,salvati00}). Based on the
$S_{\rm 9.7}$--$N_{\rm H}$ distribution of our Compton-thick AGN
sample, we should expect $\approx 6$--8 sources in this
`heavily-obscured' parameter space; however, we find only one source,
suggesting $\gtrsim 80$\% still remain unidentified in the nearby
Universe. From the sources currently identified with $S_{\rm 9.7} >
0.5$, we expect the majority of the `heavily-obscured' systems to be
optically-unidentified due to large quantities of host-galaxy dust as
well as being starburst dominated in the mid-IR. Hence, they will
still remain unidentified in high-redshift multi-wavelength
blank-field surveys. This will potentially cause a bias in our current
perceived view of the heavily-obscured AGN population and their place
in galaxy evolution models.

As we have shown, the Si-absorption features in many edge-on
Compton-thick AGN appear to be unassociated with the dusty torus
surrounding the central engine; we therefore cannot use these sources
to constrain theoretical models for the AGN infrared
emission. However, based on our results those Compton-thick AGN hosted
in face-on galaxies appear relatively unobscured in the mid-IR with
$S_{\rm 9.7} \sim 0$--0.3. Such AGN are unlikely to have significant
host-galaxy dust extinction along the line-of-sight; hence, any
observed Si-absorption will be intrinsic to the torus. For a sensible
(physical) range of AGN torus parameters (see Table 1 of Fritz et
al. 2006; Table 1 of Schartmann et al. 2008), clumpy torus models
(e.g., Nenkova et al. 2002; Schartmann et al. 2008), where-by the dust
is randomly distributed in discrete dust clouds, predict similarly
weak Si-absorption ($S_{\rm 9.7} < 0.3$) for optically-thick edge-on
($i=90^o$) tori. By contrast, uniform torus models (e.g., Pier \&
Krolik 1992; Fritz et al. 2006) predict large silicate features
($S_{\rm 9.7} \gg 1$) for similar input parameters. Indeed, for a
Compton-thick system, uniform torus models reproduce $S_{\rm 9.7} \sim
0$ only when invoking a steep dust-density gradient across an
extremely compact ($R_{max}/R_{min} \ll 30$) and somewhat unphysical
torus. Therefore, under the paradigm of a clumpy torus, we suggest
that the additional Si-absorption measured in those Compton-thick AGN
which are hosted in edge-on host-galaxies becomes entirely consistent
with extinction arising from dust which is distributed at very large
radii ($\gg$~parsecs) within the main host galaxy. In light of these
findings, in Fig. 3 we present a schematic cartoon of the effect of
the host-galaxy inclination and morphology on the observed mid-IR
Si-absorption feature.

\section{Summary \& Conclusions} \label{sec:summary}

We have investigated mid-infrared (mid-IR) dust extinction in all hard
X-ray ($E > 10$~keV) detected, {\it bona-fide} Compton-thick ($N_{\rm
  H} \gtrsim 1.5 \times 10^{24} \pcmsq$) AGN in the nearby Universe to
constrain the dominant source of attenuating dust in heavily obscured
AGN and build a more complete understanding of Compton-thick AGN at
mid-IR wavelengths. We measured Si-absorption features at 9.7$\mu m$
in archival low-resolution ($R \sim 57$--127) mid-IR {\it Spitzer}-IRS
spectroscopy to infer the attenuating dust column in these highly gas
absorbed systems and determine the average mid-IR properties of
Compton-thick AGN. We found that the mid-IR SEDs of Compton-thick AGN
are diverse, and only in a minority of sources (9/20) do we find
strong Si-absorption features ($S_{\rm 9.7} \gtrsim 0.5$). In turn, we
suggest that sample selection based solely on the mid-IR absorption
features (for $S_{\rm 9.7} \gtrsim 1$) may miss $\gtrsim 80$\% of
Compton-thick AGN. From analysis of the average mid-IR spectrum, we
find that Compton-thick AGN are characterized by roughly equal (40:60
ratio) quantities of starburst and AGN components which is only mildly
dust obscured ($S_{\rm 9.7} \sim 0.36 \pm 0.04$; $A_{\rm V} \approx
3.9$--4.9), i.e., {\it average} dust-extinction levels which are
consistent with those which are observed in host galaxy star-forming
regions. Furthermore, the high-levels of star-formation observed in
these Compton-thick AGN appears to be a feature specific to the most
heavily-obscured AGN in the nearby Universe. If such a trend continues
towards higher redshifts, then infrared photometric and X-ray surveys
which focus on the detection of powerlaw components to infer AGN
activity within a source (e.g., \citealt{daddi07,donley07,donley08})
will be partially biased against the detection of a significant
proportion of the Compton-thick AGN population, and hence, may still
miss the majority of the most heavily obscured AGN.

We compared $N_{\rm H}$ and $S_{\rm 9.7}$ in Compton-thick AGN, and
find no significant correlation between these two measures of apparent
obscuration. We find that the most heavily attenuated sources ($S_{\rm
  9.7} \gtrsim 0.5$) appear to be hosted in galaxies with visible
dust-lanes, disturbed morphologies and/or galaxies which are
highly-inclined along the line-of-sight. By contrast, we show that
sources hosted in face-on galaxies present only weak dust extinction
features ($S_{\rm 9.7} \sim 0$--0.3). A similarly narrow-range in
Si-absorption is predicted by those theoretical torus models which
invoke clumpy distributions for the obscuring material. We suggest
that the deeper silicate features ($S_{\rm 9.7} \gtrsim 0.5$), which
are observed in only a minority of Compton-thick AGN ($\approx 40$\%),
arise from intervening dust at much larger scales than predicted for a
torus (i.e., from within the host galaxy). When combined with previous
investigations, this provides further evidence that the obscuring cool
dust, which dominates the extinction seen at mid-IR (and optical)
wavelengths, may not necessarily be co-spatial with a gas-rich central
torus.

\acknowledgments

The authors would like to thank the anonymous referee for their timely
and considered report which has allowed us to clarify and improve
several aspects of the manuscript. We are also thankful to G. Risaliti
for helpful discussions. This work was partially supported by NASA
grants 13637399 and AR8-9017X. DMA acknowledges funding from the
Science and Technologies Funding Council. FEB acknowledges support
from Basal-CATA (PFB-06/2007) and CONICYT-Chile (FONDECYT 1101024 and
ALMA-CONICYT 31100004). This work is based on observations made with
the {\it Spitzer} Space Telescope and has made use of the NASA/IPAC
Infrared Science Archive which are operated by the Jet Propulsion
Laboratories, Californian Institute of Technology under contract with
NASA.  {\it Facilities:} \facility{Spitzer (IRS)}.

\bibliography{bibtex1}

\end{document}